\documentclass[twocolumn]{aastex61}
\submitjournal{ApJ}
\shorttitle{Off-limb continuum flare emission}
\shortauthors{Heinzel et al.}

\begin{document}

\title{On the nature of off-limb flare continuum sources detected by {\it SDO}/HMI}
%\author{P. Heinzel\altaffilmark{1}, L. Kleint\altaffilmark{2}, J. Ka\v{s}parov\'{a}\altaffilmark{1}, S. Krucker\altaffilmark{2}}

\author{P. Heinzel}
\affiliation{Astronomical Institute, Czech Academy of Sciences, 25165 Ond\v{r}ejov, Czech Republic}

\author{L. Kleint}
\affiliation{University of Applied Sciences and Arts Northwestern Switzerland, Bahnhofstrasse 6, 5210 Windisch, Switzerland}

\author{J. Ka\v{s}parov\'{a}}
\affiliation{Astronomical Institute, Czech Academy of Sciences, 25165 Ond\v{r}ejov, Czech Republic}

\author{S. Krucker}
\affiliation{University of Applied Sciences and Arts Northwestern Switzerland, Bahnhofstrasse 6, 5210 Windisch, Switzerland}

\correspondingauthor{P. Heinzel}
\email{pheinzel@asu.cas.cz}

%\altaffiltext{1}{Astronomical Institute, Czech Academy of Sciences, 25165 Ond\v{r}ejov, Czech Republic}
%\altaffiltext{2}{University of Applied Sciences and Arts Northwestern Switzerland, Bahnhofstrasse 6, 5210 Windisch, Switzerland}  

\begin{abstract}
The Helioseismic and Magnetic Imager onboard the {\it Solar Dynamics Observatory} has provided unique observations of off-limb flare emission. 
White-light (WL) continuum enhancements were detected in the "continuum" channel of the Fe 6173 \AA \, line
during the impulsive phase of the observed flares. In this paper we aim to determine
which radiation mechanism is responsible for such an enhancement being seen above the limb, at chromospheric heights around or below
1000 km. Using a simple analytical approach, we compare two candidate mechanisms, the hydrogen 
recombination continuum (Paschen) and the Thomson continuum due to scattering of disk radiation on flare electrons.
Both mechanisms depend on the electron density, which is typically enhanced during the impulsive phase of a flare
as the result of collisional ionization (both thermal and also non-thermal due to
electron beams). We conclude that for electron densities higher than $10^{12}$ cm$^{-3}$, the Paschen recombination
continuum significantly dominates the Thomson scattering continuum and there is some contribution from the hydrogen free-free emission. 
This is further supported by detailed radiation-hydrodynamical (RHD)
simulations of the flare chromosphere heated by the electron beams. We use the RHD code {\em FLARIX} to compute the temporal evolution
of the flare heating in a semi-circular loop. The synthesized continuum structure
above the limb resembles the off-limb flare structures detected by HMI, namely their height above the limb, as well
as the radiation intensity. These results are consistent with recent findings
related to hydrogen Balmer continuum enhancements, which were clearly detected in disk flares by the {\it IRIS}
near-ultraviolet spectrometer.

\end{abstract}

\keywords{Sun: flares -- Sun: continuum radiation}
%\maketitle

\section{Introduction}

Off-limb observations of solar flares are rather rare but they can provide an important constraint on the height
variation of different flare emissions. Occasionally, off-limb flares have been detected in the H$\alpha$ line
and some other lines, but the so-called white-light flares (WLF) observed on the disk in the visible light have never been
seen above the limb from the ground. Only recently has the full-disk monitoring of the Sun by {\it SDO} allowed such
detections through the Helioseismic and Magnetic Imager (HMI) instrument, thanks to observing conditions in space. 
What we see above the limb at the outermost wavelength channels 
of the Fe I 6173  \AA \, line is believed to be continuum emission unaffected by the line itself. This channel is often used to
detect the white-light (WL) continuum in disk flares and various such observations have been reported in the literature. However,
above the limb, only a few cases have been analyzed (\cite{Oliveros2012}, \cite{Krucker2015}) that challenge our understanding of the continuum 
formation in flares.

WL continuum enhancements seen on the disk inside the flare ribbons (e.g. \cite{Kuhar2016}) are normally interpreted as
being due to
either hydrogen recombination emission in the Paschen continuum (below the series limit at 8203 \AA), or photospheric
(i.e. H$^{-}$) continuum enhancement, or a combination of both. However, the continuum enhancement on the disk
relative to the photospheric brightness itself (i.e. the contrast) is usually low in the visible range
(typically below 10\%) and
this makes it very difficult to disentangle the two emission mechanisms (see, e.g. \cite{Kerr2014}). Note that the Paschen recombination
continuum intensity is related to the Balmer continuum intensity and the latter was recently detected by {\it IRIS} in its
near-ultraviolet (NUV)
channel (\cite{Heinzel2014}, \cite{Kleint2016}). There is no suitable flare observation yet above the limb by {\it IRIS}
that 
would show the Balmer continuum. However, contrary to disk observations (both ground-based and using HMI), the off-limb
HMI WL continuum detection can reveal important height distributions of the emissivity and thus can help to disentangle
the various possible mechanisms. 

Another HMI off-limb observation of a WL emission comes from high-lying (33 Mm) flare loops seen during the
gradual phase of a flare \citep{Saint-Hilaire2014}, where significant degree of linear polarization was also detected. This emission was then interpreted as being due to the Thomson scattering on free electrons inside the loops.

In this paper we investigate the nature of the off-limb emission at flare footpoints, using simple analytical relations for
WL emissivities and using also using the results of radiation-hydrodynamical (RHD) simulations of flare
heating by electron beams. For the latter we use the RHD code {\it FLARIX}. We compare the synthetic continuum intensity
with that detected by HMI and discuss constraints set by HMI off-limb observations on flare-heating models.

    \begin{figure*}
    \centering
%    \figurenum{1}              
        \includegraphics[height=5.5cm, angle=0]{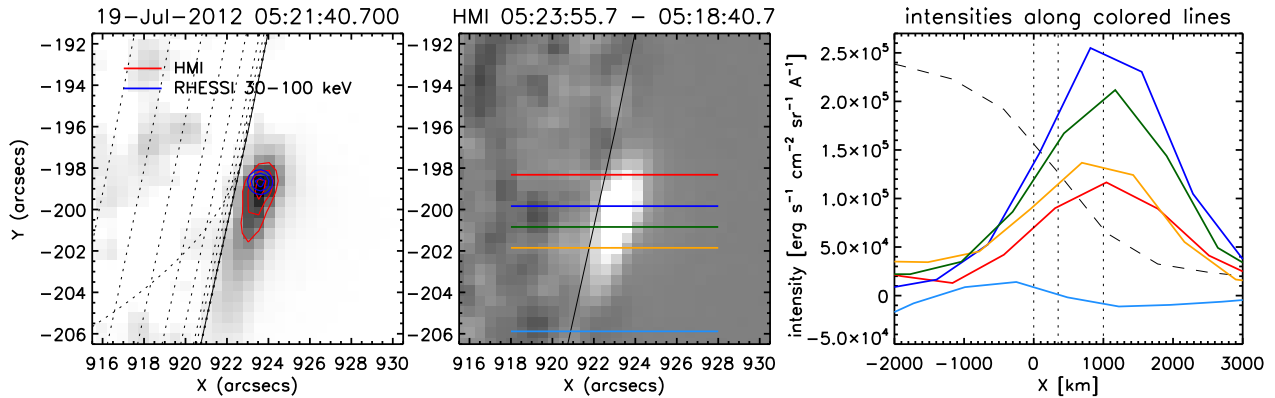}
    \caption{Co-spatial white-light and hard X-ray flare footpoints seen above the solar limb for the
    20120719 flare. Note that all curves go to zero in the corona above the flare, due to the
                  subtraction of the pre-flare emission. The intensity thus represents pure flare continuum
                  emission. The three vertical dashed lines on the rightmost panel represent the zero height at
                  $\tau_{5000}=1$, the photospheric limb at height 350 km, and chromospheric height at 1000 km,
                  respectively.}
               \label{flare20120719}
    \end{figure*}    

    \begin{figure*}
    \centering
%    \figurenum{1}              
        \includegraphics[height=5.5cm, angle=0]{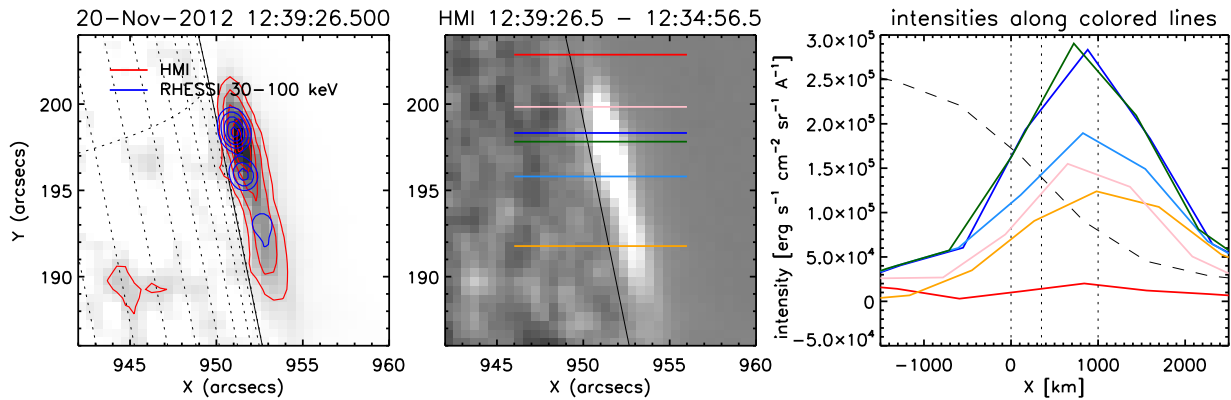}
    \caption{Same as a) but for the 20121120 limb flare}
               \label{flare20121120}
    \end{figure*}    

\section{{\it SDO}/HMI Off-limb Flare Observations}

To obtain the observed intensities of off-limb sources, we analyze 
hmi.Ic\_45s {\it SDO}/HMI data.
% SDO reference
We investigate two 
limb flares: the M7.7 flare on 2012 July 19 and the M1.7 flare on 
2012 November 20. \cite{Krucker2015} determined the heights of their WL 
emission above $\tau_{5000}=1$ to be 824 $\pm$ 70 km and 799 $\pm$ 70 km, respectively. The left 
panel of Fig. 1 is adapted from \cite{Krucker2015} and shows a 
negative difference image for the WL emission. The red contours trace the 
HMI emission and blue contours show hard X-ray emission at 30-100 
keV, which is co-spatial to the continuum emission. The middle panel 
shows a difference image at the time of maximum enhancement. The limb is 
drawn by the SSW routine plot\_map.pro and is based on a calculation. Intensity cuts along 
the five horizontal lines are shown in the right panel. We converted from 
DN/s to absolute units by taking the HMI disk-center intensity and 
setting it to the continuum value at 6173 \AA \, of 0.315 $\times$ 10$^7$ 
erg s$^{-1}$ cm$^{-2}$ sr$^{-1}$ \AA$^{-1}$ from Brault \& Neckel's atlas \citep{neckel1994}. 
Here we determined the limb as the inflexion point of the pre-flare limb 
darkening intensity, which differs from the calculated limb of plot\_map 
by about one pixel ($0.\!\arcsec 5$) and we assigned it the height 350 km 
(the $x$-axis in the right panel). The maximum observed enhancements are 
$3\ \times\ 10^5\ \mbox{erg s}^{-1}\ \mbox{cm}^{-2}\ \mbox{sr}^{-1}\ \AA^{-1}$, and the spatial and temporal variations are 
significant. Our height estimate of the peak emission from Figures 1 and 2 is in the range of 800 - 1100 km
above the level of $\tau_{5000}=1$.

\section{Mechanisms of the WL Continuum Emission}

As stated in the introduction, the WL enhancement on the disk consists of the expected photospheric contribution due to
H$^{-}$ and the hydrogen recombination continuum, which is the Paschen continuum around the 6173 \AA \, HMI line. Above the limb 
at chromospheric heights, we expect that HMI observations should reveal only the Paschen component.
However, some HMI off-limb observations have been interpreted in terms of Thomson scattering and we therefore devote
this section to an analytical estimation of the relative importance of the Paschen vs. Thomson components. Moreover, we 
estimate the importance of the hydrogen free-free emission above the limb.

\subsection{Paschen Continuum Emission Due to Hydrogen Recombination}

Hydrogen in the flaring chromosphere is highly ionized due to a strong temperature increase (thermal collisional ionization), but also due to 
collisions of hydrogen with non-thermal electrons from the beam. Protons then capture free thermal electrons (note that the non-thermal ionization
of hydrogen produces electrons that are finally thermalized) and create neutral hydrogen atoms, a process called hydrogen
recombination. It has been demonstrated that the possible capture of non-thermal electrons from the beam is quite negligible.

The Paschen continuum emissivity is then expressed as \citep{HM2015}

\begin{equation}
\eta_{\nu}^{i} = n_{\rm e} n_{\rm p} F_i(\nu, T) \, ,
\end{equation}
where $i$=3 (Paschen continuum), $n_{\rm e}$ and $n_{\rm p}$ are the electron and proton densities, respectively, 
$\nu$ is the frequency of the observed emission, and $T$ is the kinetic temperature of the source. The function $F_i$ has the
form

\begin{eqnarray}
F_i(\nu, T)  & = & 1.166 \times 10^{14} g_{\rm{II}}(i,\nu) T^{-3/2} B_{\nu}(T) \times \nonumber \\
  &  & {\rm e}^{h\nu_i/kT} (1 - {\rm e}^{-h\nu/kT} )/ (i \nu)^3 \, .
\end{eqnarray}
Here, $\nu_{i}$ is the frequency at the continuum head, 
$g_{\rm II}$ is the Gaunt factor, $B_{\nu}(T)$ is the Planck function, and $h$ and $k$ are the Planck and Boltzmann constants,
respectively. For optically thin plasma, we multiply this emissivity by the line-of-sight geometrical extension of the
source to obtain the specific intensity of the continuum radiation $I(\nu)$.

% Table  T vs. F_i
%  6000 K    5.283-12
%10000 K    3.609-12
%15000 K    2.382-12
%30000 K    1.021-12
%50000 K    5.125-13

\subsection{Thomson Scattering}

Here we compute the continuum intensity around the 6173 \AA \, line due to Thomson scattering on flare-loop electrons. The scattering
is assumed to be coherent and isotropic. The emissivity is given by the standard formula (e.g. \cite{HM2015})

\begin{equation}
\eta_{\nu} = n_{\rm e} \sigma_{\rm{T}} J(\nu) \, ,
\end{equation}
where $\sigma_{\rm{T}}$ = 6.65 $\times$ 10$^{-25}$  cm$^2$ is the absorption cross section for Thomson scattering 
and $J(\nu)$ is the diluted photospheric WL radiation. The latter is obtained from the disk-center continuum
intensity, multiplied by the dilution factor for chromospheric heights. A purely geometrical dilution factor would be
around 0.5 for low heights, up to 1000 km. However, here we take into account the center-to-limb variations
of the photospheric continuum intensity and thus the effective dilution factor is 0.354 at 6174 \AA \, (\cite{Jejcic2009}).
The disk-center continuum intensity at 6173 \AA \, is
$3.96\ \times\ 10^{-5}\ \mbox{erg s}^{-1}\ \mbox{cm}^{-2}\ \mbox{sr}^{-1}\ \mbox{Hz}^{-1}$, as linearly
interpolated from Allen's tables \citep{Cox2000}. This, multiplied by $0.354$ gives a value of
$J(\nu)=1.40\ \times\ 10^{-5}\ \mbox{erg s}^{-1}\ \mbox{cm}^{-2}\ \mbox{sr}^{-1}\ \mbox{Hz}^{-1}$, which is used in our modeling.
In reality, the scattering is strictly coherent only in the electrons' rest-frames, but due to their thermal
motions the scattered photons are Doppler-shifted. In the case of scattering of the photospheric line radiation, 
this leads to a smearing of the scattered line profile, which is well known in the coronal WL spectra. In our case,
the non-coherent scattering of the 6173 \AA \, line radiation could somewhat affect our results, but using the 
above formula we get an upper limit for the continuum intensity due to Thomson scattering. 
The assumption of isotropic scattering is standard in atmospheric modeling \citep{HM2015}.

\subsection{Relative Importance of the Paschen and Thomson Continuum above the Limb}

From the above equations we see that the ratio of the intensities of the Paschen and Thomson continuum will be proportional to the
electron density under the assumption that $n_{\rm e} = n_{\rm p}$, which is well satisfied in the chromosphere where the helium
contribution to the electron density is small:

\begin{equation}
\frac{I^{\rm Pa}}{I^{\rm Th}} = n_{\rm e} \frac{F_i}{\sigma_{\rm T} J(\nu)} \, .
\end{equation}
We show this ratio for a range of electron densities and for various plasma temperatures (Figure 3).
While at lower electron densities, say, below 10$^{12}$ cm$^{-3}$, the Thomson continuum starts to dominate over the Paschen continuum, for
the higher densities usually met in the flaring chromosphere, the Paschen continuum dominates. The ratio decreases with increasing
kinetic temperature, which is due to the nature of the recombination process. Since the electron densities in stronger flares can reach
10$^{13}-10^{14}\ \mbox{cm}^{-3}$ \citep[e.g.][]{Avrett1986}, we immediately see that any off-limb emission must be quite dominated by the
Paschen continuum. In the next section we will confirm this estimate using and RHD simulation of the electron-beam
heated atmosphere.

    \begin{figure}
    \centering
%    \figurenum{1}              
        \includegraphics[height=7cm, angle=0]{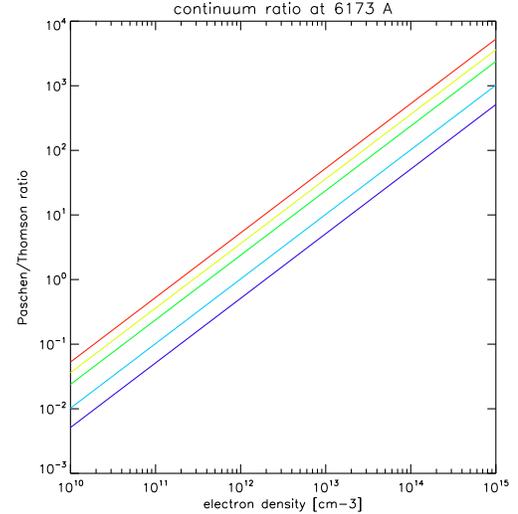}
    \caption{Theoretical ratio of the Paschen to Thomson continuum intensity at 6173 \AA, as a function of the
                  electron density. The colors correspond to different temperatures: red - 6000 K, yellow - 10,000 K,
                  green - 15,000 K, blue - 30,000 K, magenta - 50,000 K.}
               \label{ratio}
    \end{figure}

\subsection{Hydrogen Free-Free Continuum Emission}
So far we  have discussed the most explored mechanisms of the WL emission in solar flares. However, in the optical range and for longer wavelengths,
hydrogen free-free continuum emission may also contribute, namely at higher temperatures. One can easily express the ratio between free-free and
Paschen free--bound continuum intensity as

\begin{equation}
\frac{I^{\rm ff}}{I^{\rm Pa}} = 8.55 \times 10^{-5} T  {\rm e}^{h\nu_i/kT} \, ,
\end{equation}
using the expressions for emissivities from \cite{HM2015} and assuming that the Gaunt factors are around unity. 
This immediately shows that such a ratio substantially increases with increasing 
temperature, being equal to 0.15 for $T=10,000$ K, 0.71 for 20,000 K, and 1.43 for 30,000 K. Therefore, between
$T=20,000$ K and 30,000 K, the
free-free continuum emission starts to dominate over the Paschen continuum.

\section{RHD Simulation with the {\em FLARIX} Code}

In this section we present our RHD simulation of the evolving flare chromosphere and use it
to visualize the continuum source above the limb. The RHD code {\em FLARIX} \citep[see, e.g.][]{Varady2010,Heinzel2016} was
run for a specific model of the electron-beam heating inside a semi-circular loop extending from the photosphere to
coronal heights. The initial model was the VAL-C type atmosphere \citep{VAL1981} with an enhanced pressure/density.
The radiation losses of this starting atmosphere were
balanced by a constant heating term to assure initial hydrostatic equilibrium. The electron beam has a triangular
temporal variation of the energy flux lasting 10 s, with a peak flux  of 0.75 $\times$ 10$^{11}$ erg s$^{-1}$ cm$^{-2}$ at time $t$=5 s.
The spectral index of the beam was $\delta=4$. Optically thick radiation losses and hydrogen non-thermal collisional
rates have been computed at chromospheric heights. To synthesize the off-limb continuum structure, we use a snapshot
at time $t$=4 s from the whole {\em FLARIX} simulation. At this time, just before the peak of the energy deposit, the middle
chromosphere is already highly ionized and thus a further increase of the flux practically does not affect our results. The peak flux
derived from an analysis of RHESSI spectra of these flares is higher, but that seems to affect the atmospheric electron
density only marginally. Moreover, electron-beam fluxes exceeding 10$^{11}$ erg s$^{-1}$ cm$^{-2}$ could produce a significant return current
in the chromosphere and this is not yet properly treated in the available RHD codes. In Figures 4 and 5 we show the height variations of the
kinetic temperature and electron density at this time, respectively. 

    \begin{figure}
    \centering
%    \figurenum{1}              
        \includegraphics[height=7cm, angle=0]{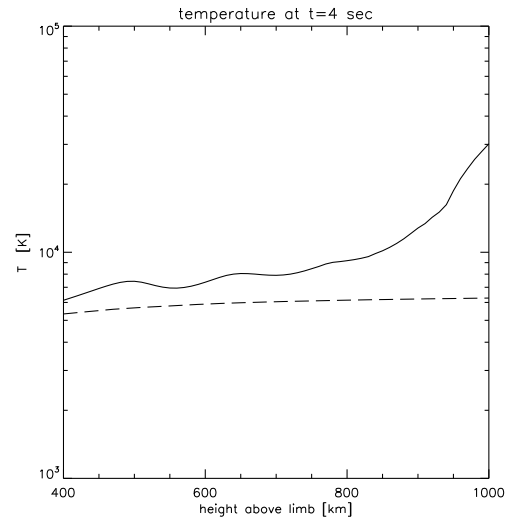}
    \caption{Snapshot of the height variation of the kinetic temperature from the {\em FLARIX} simulation at time $t$=4 s.
    For comparison, we also show the temperature structure of a pre-flare VAL3C atmosphere with an enhanced plasma density
    (dashed line). The zero height (the limb) is located at 350 km above the level of $\tau_{5000}=1$, i.e. the height scale is shifted compared to Figures 1 and 2.}
               \label{temp}
    \end{figure}    
    
    \begin{figure}
    \centering
%    \figurenum{1}              
        \includegraphics[height=7cm, angle=0]{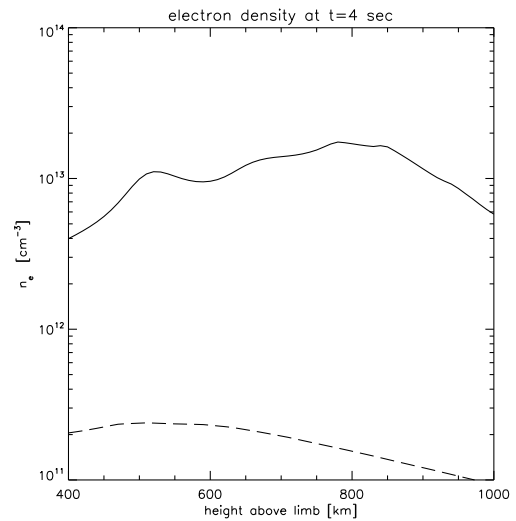}
    \caption{Snapshot of the height variations of the electron density from the {\em FLARIX} simulation at time $t$=4 s.
    For comparison, we also show the electron density of a pre-flare VAL3C atmosphere with an enhanced plasma density
    (dashed line). The zero height (the limb) is located at 350 km above the level of $\tau_{5000}=1$. }
               \label{elden}
    \end{figure}    
    
        \begin{figure}
    \centering
%    \figurenum{1}              
        \includegraphics[height=7cm, angle=0]{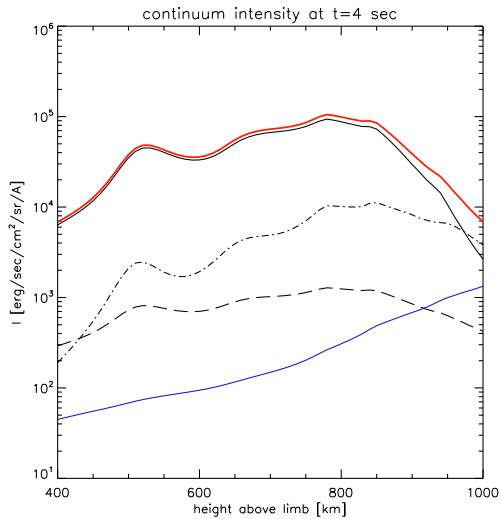}
    \caption{Vertical variations of the line-of-sight total continuum intensity for thickness 1000 km, from our {\em FLARIX} simulation at time $t$=4 s (red line). This is
    mostly dominated by the Paschen continuum (full black line), while the dotted-dashed line shows the hydrogen free-free contribution and the dashed line shows the Thomson scattering component.
    The blue line indicates the energy deposit rate due to the electron-beam precipitation.
    The zero height (the limb) is located at 350 km above the level of $\tau_{5000}=1$. 
    }
               \label{continua}
    \end{figure}    

From Figure 5 we immediately see that the electron density in the middle chromosphere reaches values above
10$^{13}$ cm$^{-3}$ which means that, according to our plot
in Figure 3, the continuum emission above the limb is fully dominated by the Paschen continuum, while the Thomson scattering is 
quite negligible. Note that the Paschen-continuum is optically thin under these conditions, so  the line-of-sight
integration of the emissivity is straightforward and gives the synthetic intensity. 

\subsection{1.5D Visualization of the Off-limb Structure}

    \begin{figure}
    \centering
%    \figurenum{1}              
        \includegraphics[height=5cm, angle=0]{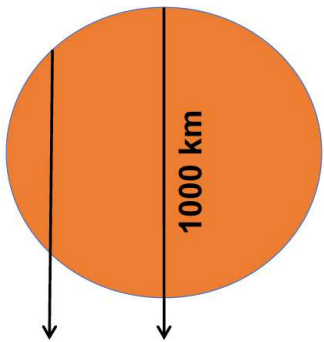}
    \caption{Horizontal cross section of the flare loop showing the two selected lines of sight along which we integrate the
    continuum emissivities. The loop diameter is 1000 km.}
               \label{cross}
    \end{figure}    

    \begin{figure}
    \centering
%    \figurenum{1}              
        \includegraphics[height=6cm, angle=0]{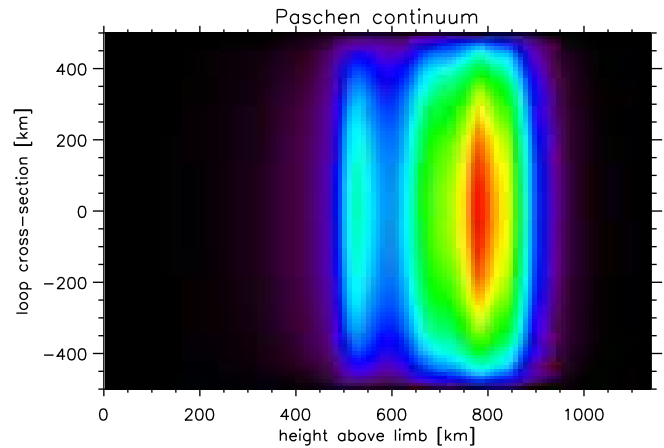}
    \caption{1.5D visualization of the off-limb structure from {\em FLARIX} simulation at time $t$=4 s.
    The zero height (the limb) is located at 350 km above the level of $\tau_{5000}=1$. }
               \label{loop}
    \end{figure}    

Assuming that the flare loop thickness is 1000 km, we integrate the continuum emissivity along the off-limb line of
sight at each atmospheric height above the limb from 0 to 1200 km. For a given height we assume a constant temperature and
electron density along the line of sight, both given by the {\em FLARIX} simulation. The resulting height-dependent continuum
intensity along the loop axis is shown in Figure~6. From this plot we already see that the off-limb
structure extends from about 450 to 950 km above the limb, which is consistent with HMI observations.
Note that the HMI brightness is displayed in Figures 1 and 2 on a scale where zero is at the level of $\tau_{5000}=1$ and thus it goes to larger heights compared
to Figures 4-6,  where zero (the limb) is at photospheric height 350 km.  Also, the intensity,
which reaches $10^5\ \mbox{erg s}^{-1}\ \mbox{cm}^{-2}\ \mbox{sr}^{-1}\ \AA^{-1}$ is consistent with that from HMI as shown in Figure 1, provided that
the actual geometrical thickness of the off-limb structure is a factor up to 3 larger than our nominal value 1000 km. This is quite
plausible because a direct inspection of the loop-system geometry from {\it STEREO} observations reveals larger line-of-sight
volumes \citep[see, e.g.~][]{Krucker2015}. Rather complex projection effects of the whole loop arcade could also be
responsible also for the wider height distribution of the continuum emission (see the rightmost panels in
Figures 1 and 2), 
compared to our single-loop simulation. In Figure 6  we also separately show
the hydrogen free-free continuum intensity that starts to dominate the Paschen continuum at temperatures above 20,000 K and
the Thomson continuum intensity that is about two orders of magnitude  weaker compared to the Paschen continuum.
For reference, we also add the curve of the energy deposit due to the electron-beam energy losses (in arbitrary units). In the low chromosphere,
the energy deposit is small compared to that at higher layers, thus the atmosphere does not differ much from the initial state.
Note that the limb in these plots is actually 
placed at a height of 350 km above the zero atmospheric height used in {\em FLARIX}, i.e. the photospheric height
where $\tau_{5000}=1$ in the quiescent atmosphere. The value of 350 km comes from Table I of \cite{Lites1983} and indicates the limb height at
wavelength 6000 \AA. This should correspond to the limb as detected by HMI.

As a next step we have 
performed a line-of-sight integration of the Paschen continuum emissivity
along the rays crossing a circular loop with a nominal diameter of 1000 km (see the top view of such loop in
Figure 7),
assuming again that the temperature and electron density in the horizontal plane are constant and equal to values
computed by {\em FLARIX} at a given atmospheric height. Since {\em FLARIX} represents a 1D modeling along
the loop axis (in the vertical direction), while the formal transfer solution is made along rays crossing our
circular loop horizontally, we call this kind of spectrum synthesis a 1.5D radiative-transfer visualization. The result is displayed
in Figure 8, where the horizontal axis shows the height above the limb and the vertical axis gives the distance from the
loop axis toward its surfaces at $\pm$ 500 km (the total thickness of the loop is 1000 km).  Figure 8 shows the off-limb distribution of the
Paschen-continuum brightness (the intensity values along the loop axis, i.e. for $y$=0, can be inferred from
Figure 6). 
From Figure 8 we see that below 400 km the continuum emission is quite negligible and the same
also applies for heights above 1000 km (where the free-free emission already dominates). 
The brightness variations along the $x$-axis are mainly due to the height variations
of the electron density computed by {\em FLARIX}. Note that the intensity is roughly proportional to $n_{\rm e}^2$. The main peak
is between 750 and 800 km above the limb, similar to HMI observations of selected structures (see Figure 1).  The less bright 
secondary peak around the height 500 km, not seen in the HMI observations (note, however, that we did not perform any convolution with the
HMI PSF), is the result of our particular simulation and different {\em FLARIX} models will likely lead  to different height distributions
of the continuum emission.

\section{Discussion and Conclusions}

Our results set quantitative constraints on the mechanisms of WL emission in solar flares, the issue frequently referred
to as the ``WLF mystery''.
Contrary to more frequent disk observations where the WL enhancement is likely due to a mixture of Paschen-continuum emission and 
enhanced photospheric H$^-$ continuum, the off-limb observations naturally eliminate the latter contribution. Our estimates, together with
detailed numerical simulations, lead to the conclusion that the observed WL enhancement detected by {\it SDO}/HMI is  dominated by the
hydrogen Paschen continuum (Brackett continuum is about 5 time weaker), with a small contribution from the free-free mechanism.
Thomson scattering on chromospheric electrons is also enhanced during
the flare, compared to a pre-flare atmosphere, but because the ratio of Paschen/Thomson emissivity scales linearly with the electron density,
the Paschen continuum significantly dominates, namely in stronger flares. However, in the case of the so-called 'post-flare' loops observed by HMI
in WL at high altitudes around 33 Mm, \cite{Saint-Hilaire2014} detected significant $Q$-polarization, as expected from pure Thomson
scattering at such altitudes and wavelengths, for electron densities below 10$^{11}$ cm$^{-3}$. 
This is consistent with our estimates presented in Figure 3. Note that in solar prominences where the electron densities are 
even lower, the WL continuum is indeed dominated by the Thomson scattering \citep{Jejcic2009}. At lower altitudes around 15 Mm, \cite{Saint-Hilaire2014}
found a dominant non-Thomson component (at densities around 10$^{12}$ cm$^{-3}$) and suggested their free-free and free-bound origin.

Our simulations, and namely 1.5D visualizations, are based on a circular loop-like model for which we took a diameter of 1000 km. 
Several such loops can be aligned along the line of sight, as {\it STEREO} observations of this flare do indicate. However, in reality these
flare loops can reach diameters as small as 100 km, as recent high-resolution observations suggest (\cite{Jing2016}). The latter authors
conclude that the whole flare loop system (they show cool 'post' flare loops, but those result from previous hot loops considered here) has
a very small filling factor compared to volumes visible at much lower spatial resolutions like those seen by HMI or RHESSI. This issue will set new
constraints on future quantitative modeling, where also the assumed electron-beam fluxes may reach even higher values.

We have also investigated the surrounding chromospheric opacity of the bound-free Paschen continuum. Using the quiet-Sun
model VAL-C we have integrated the opacity along the off-limb lines of sight at various chromospheric heights where the flare
emission is detected. The corresponding optical thicknesses of the Paschen continuum at the HMI wavelength were found to be quite small,
around 10$^{-3}$ just above the limb and decreasing upward to about 1.4 $\times 10^{-4}$ at heights around 800 km above the limb
where our simulations give the maximum of the flare continuum emission. Therefore, we do not expect 
the surrounding chromosphere to have any effect on the flare continuum visibility. In this estimate, we neglected the possible influence of spicules seen above
the limb; in actuality, any non-negligible continuum opacity would lead to an obscuration of the off-limb flare and 
also would produce a bright ring in the chromosphere surrounding the 
solar limb, which,for example, is observed in H$\alpha$, but not in WL.

Off-limb observations of flares with HMI could also provide us with information about the formation height of the iron line 6173 \AA.
It is well known that the cores of many metallic (iron) lines are enhanced during flares \citep[see, e.g., the recent paper by][]{Kleint2017}
and it would be very interesting to see at which heights they are formed during the flare - these are normally photospheric
lines when formed in an undisturbed atmosphere. However, it is assumed that the outermost wavelength channels of HMI do not
contain line emission and are thus solely representative of the WL continuum emission. 
Our inspection of HMI off-limb spectra indicates that the intensity enhancement is about the same at all wavelengths.
But the Thomson scattering of the 6173 \AA \, line radiation can result in a lowering of the WL intensity, compared to our model in which
the pure continuum scattering is considered for simplicity. 

Finally, it would be quite interesting to obtain limb observations of flares by {\it IRIS}, in order to detect an off-limb Balmer-continuum emission, 
similar to that found on the disk by \cite{Heinzel2014} and \cite{Kleint2016}. This, together with the HMI observations discussed in this
paper, could provide a unique constraint on the radiation mechanisms of white-light flares. Off-limb emission in the hydrogen Balmer continuum
would observationally constrain the chromospheric losses in this continuum. Such losses have been shown
to be dominant in the region of the electron-beam energy deposit and thus they play a crucial role in the energy balance \citep{Heinzel2016}.
Based on the results of the existing HMI off-limb and {\it IRIS} on-disk observations of the continuum emission, the issue of WLFs
becomes less ``mysterious'', although the true mechanism behind the photospheric (on-disk) WL enhancement is still not well understood.

% Mention relaxation times for recombination - Paschen light curves can be delayed behind more instantaneous HXR bursts
% as found for the case of Balmer recombination continuum detected in the disk flare (Heinzel \& Kleint 2014).

\acknowledgments{
This work was supported by the grant No. 16-18495S of the Czech Funding Agency and by ASI ASCR project RVO:67985815. 
The authors acknowledge support by the International Space Science Institute (ISSI), and namely through the team of
H. Tian, who deal with chromospheric flares. Comments and suggestions of the referee are also highly acknowledged.
}

%\begin{thebibliography}
%\end{thebibliography}

\bibliographystyle{aasjournal}
\bibliography{heinzel}

\end{document}